\documentclass[showpacs,twocolumn,preprintnumbers,amsmath,amssymb,amsfonts,prl,floatfix]{revtex4}

\usepackage{graphicx, color,rotating} % Include figure files
\usepackage{dcolumn} % Align table columns on decimal point
\usepackage{bm} % bold math

\begin{document}

% Title Page
\title{Spin-dependent Cooper Pair Phase and 
Pure Spin Supercurrents \\in Strongly Polarized Ferromagnets}

\author{R.~Grein}
\affiliation{Institut f\"ur Theoretische Festk\"orperphysik
and DFG-Center for Functional Nanostructures,
Universit\"at Karlsruhe D-76128 Karlsruhe Germany}
%\homepage{http://www.tfp.uni-karlsruhe.de}
%\homepage{http://www.cfn.uni-karlsruhe.de}
\author{M.~Eschrig}
\affiliation{Institut f\"ur Theoretische Festk\"orperphysik
and DFG-Center for Functional Nanostructures,
Universit\"at Karlsruhe D-76128 Karlsruhe Germany}
\author{G.~Metalidis}
\affiliation{Institut f\"ur Theoretische Festk\"orperphysik
and DFG-Center for Functional Nanostructures,
Universit\"at Karlsruhe D-76128 Karlsruhe Germany}
\author{Gerd~Sch\"on}
\affiliation{Institut f\"ur Theoretische Festk\"orperphysik
and DFG-Center for Functional Nanostructures,
Universit\"at Karlsruhe D-76128 Karlsruhe Germany}
\date{\today}

\begin{abstract}
We study heterostructures of singlet superconductors (SC) and strongly
spin-polarized ferromagnets (sFM) and show that a relative phase 
arises
between the superconducting proximity amplitudes in the two ferromagnetic
spin bands.
We find a tunable
pure spin supercurrent in a sFM contacted with only one SC electrode.
We show that
Josephson junctions are most effective for a spin polarization $P\sim 0.3$,
and that critical currents for positive and negative bias differ for
a high transmission Josephson junction,
due to a relative phase
between single and double pair transmission.
\end{abstract}

\maketitle

\renewcommand{\phi}{\varphi}
\newcommand{\eps}{\varepsilon}
\newcommand{\ud}{\uparrow,\downarrow}
\renewcommand{\u}{\uparrow}
\renewcommand{\d}{\downarrow}
\newcommand{\ket}[1]{| {#1}\rangle}
\newcommand{\bra}[1]{\langle {#1}|}
\newcommand{\barlambda}{{\lambda \!\!\!^{-}\,\!}}
\newcommand{\blFe}{{\lambda \!\!\!^{-}\,\!}_{\mathrm{F}1}}
\newcommand{\blFeta}{{\lambda \!\!\!^{-}\,\!}_{\mathrm{F}\eta}}
\newcommand{\blF}{{\lambda \!\!\!^{-}\,\!}_{\mathrm{F}}}
\newcommand{\blJ}{{\lambda \!\!\!^{-}\,\!}_{J}}
\newcommand{\EF}{E_{\mathrm{F}}}
\newcommand{\vpfe}{\vec{p}_{\mathrm{F}1}}
\newcommand{\vpfz}{\vec{p}_{\mathrm{F}2}}
\newcommand{\vpfd}{\vec{p}_{\mathrm{F}3}}
\newcommand{\vpfeta}{\vec{p}_{\mathrm{F}\eta }}
\newcommand{\vvfe}{\vec{v}_{\mathrm{F}1}}
\newcommand{\vvfz}{\vec{v}_{\mathrm{F}2}}
\newcommand{\vvfd}{\vec{v}_{\mathrm{F}3}}
\newcommand{\vvfzd}{\vec{v}_{\mathrm{F}2,3}}
\newcommand{\vvfeta}{\vec{v}_{\mathrm{F}\eta }}
\newcommand{\pfe}{p_{\mathrm{F}1}}
\newcommand{\pfz}{p_{\mathrm{F}2}}
\newcommand{\pfd}{p_{\mathrm{F}3}}
\newcommand{\pfeta}{p_{\mathrm{F}\eta }}
\newcommand{\vfe}{v_{\mathrm{F}1}}
\newcommand{\Nfe}{N_{\mathrm{F}1}}
\newcommand{\vfz}{v_{\mathrm{F}2}}
\newcommand{\vfd}{v_{\mathrm{F}3}}
\newcommand{\vfzd}{v_{\mathrm{F}2,3}}
\newcommand{\vfeta}{v_{\mathrm{F}eta}}
\newcommand{\JFM}{J_{\mathrm{FM}}}
\newcommand{\vJFM}{\vec{J}_{\mathrm{FM}}}
\newcommand{\JI}{J_{\mathrm{I}}}
\newcommand{\vJI}{\vec{J}_{\mathrm{I}}}
\newcommand{\VI}{V_{\mathrm{I}}}

%\textit{INTRODUCTION} -
Superconductor(SC)/ferromagnet(FM)-hybrid structures have triggered
considerable research activities in recent years \cite{EfetovRev,Buzdin,Eschrig03,volkov,kopu,Keizer,EschrigLTP,houzet1,Nazarov,Tanaka,EschrigHM}.
In particular, FM Josephson junctions are promising spintronics devices as
they allow for tuning the critical current via the electron spin.
However, due to the competition between the uniform spin alignment in the FM and
spin-singlet pairing in the SC, singlet superconducting correlations
decay in the FM on a much shorter length scale
than in a normal metal \cite{buzdin82}.
Although this results in a rapidly decaying Josephson current for long junctions,
the proximity effect leads to interesting physics in short and/or weakly
polarized junctions, e.g., oscillations of the supercurrent as a function of
the thickness of the interlayer
that can give rise to $\pi$-junction behavior \cite{ryazanov,buzdin82}.
Recently however, in contradiction with these
expectations, long-range supercurrents have been reported through strongly
spin-polarized materials \cite{Keizer}. Theoretical calculations have shown that for
strongly polarized ferromagnets (sFM) spin scattering at SC/FM interfaces
\cite{tokuyasu88} leads to a
transformation of singlet correlations in the SC into triplet correlations
\cite{Eschrig03}
(the `triplet reservoirs' of Ref. \cite{Nazarov}), that can carry
a long-range supercurrent through the
sFM \cite{Eschrig03,EschrigLTP,houzet1,Tanaka,Nazarov,EschrigHM}.

So far, transport calculations in SC/FM hybrids have mostly
concentrated on either fully polarized FMs, so-called half metals (HM),
or on the opposite limit of weakly polarized systems.
However, most FMs have an intermediate exchange
splitting of the energy bands of the order of 0.2-0.8 times the Fermi
energy $\EF $. For this intermediate range, one could naively expect a behavior
similar to two shunted half metallic junctions.
We will show, using a microscopic interface model, that this picture
is inadequate, and point out the crucial role played by the interfaces
in coupling the sFM spin bands.

\begin{figure}[b]
\includegraphics[width = 0.41\columnwidth]{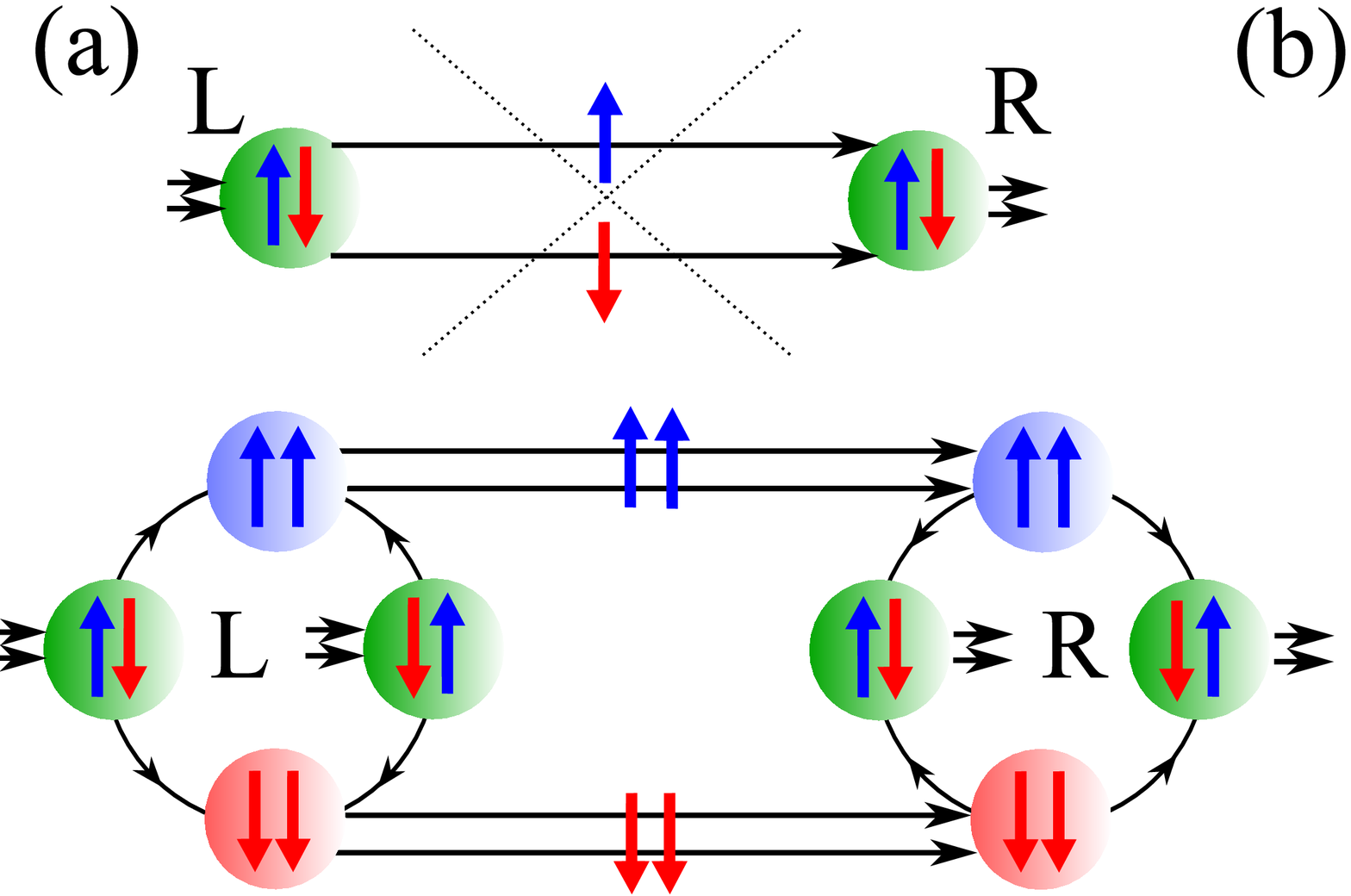}
\includegraphics[width = 0.55\columnwidth]{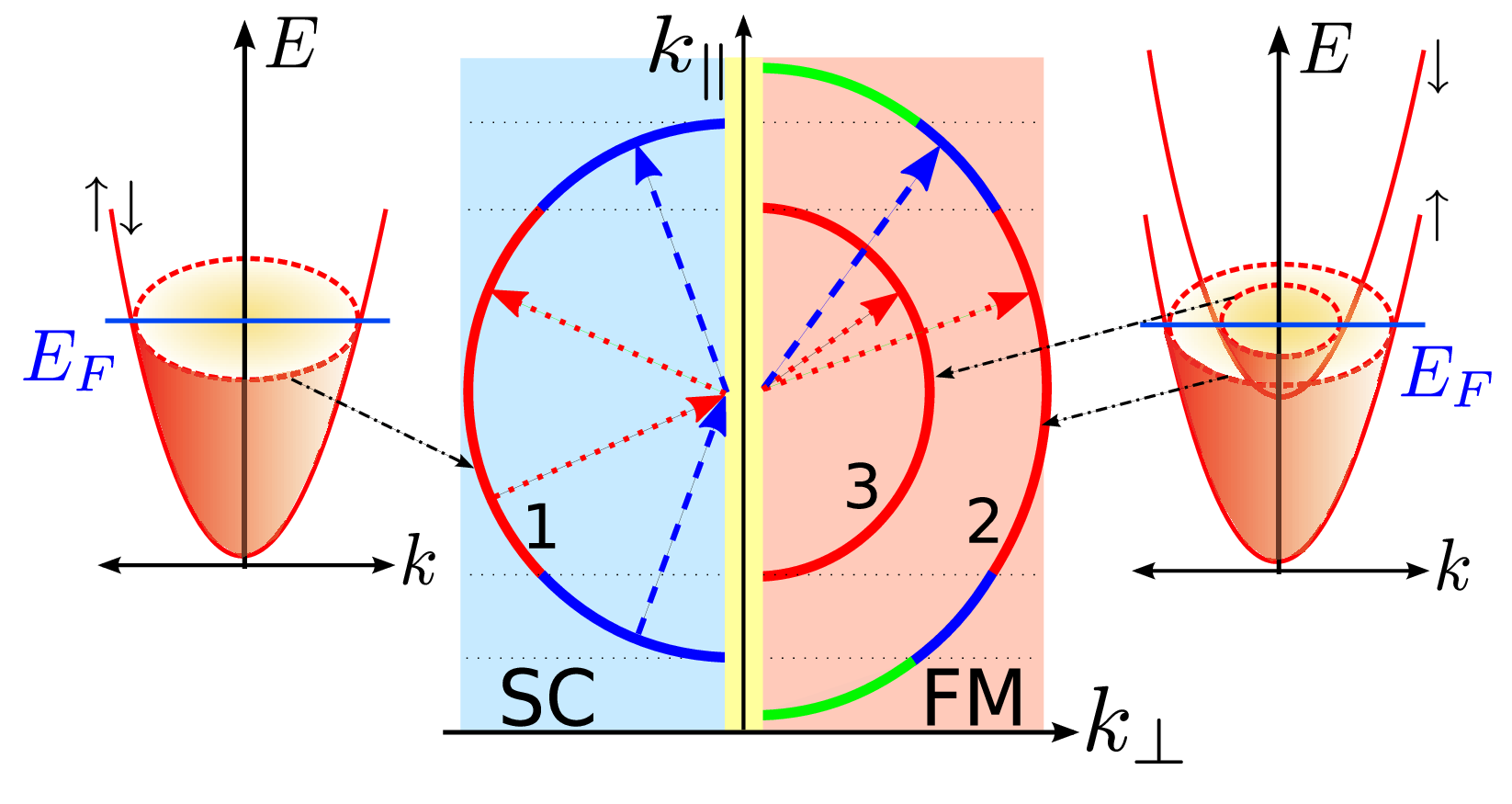}
\caption{\label{fig1}
(color online) (a) 
The coherent transfer of singlet pairs via a sFM (top) is not possible.
However, the ``crossed'' pair transmissions process (bottom) is possible
and leads to intriguing effects in high transmission junctions.
(b) SC/sFM interface, showing the Fermi surfaces on either side (thick lines).
Assuming momentum conservation parallel to the interface ($\vec{k}_{||}$),
a quasiparticle incident from the SC can either
scatter into two (dotted arrows), or into
only one (dashed arrows) spin band of the FM.
%Total reflection is possible for large $|\vec{k}_{||}|$ (green lines in FM).
}
\end{figure}

In this work, we study Josephson junctions
with a strongly polarized interlayer, and find fundamental differences
compared to both half metallic and weakly polarized interlayers. In
particular, we see that although correlations between $\u$-
and $\d$-electrons are suppressed due to the strong exchange field,
spin-active interfaces generate interactions between
long-range triplet supercurrents in the two spin bands.
We find that the long-range critical Josephson current
varies non-monotonously with spin polarization $P$, showing a maximum around
$P=0.3$.
Furthermore,
additional phases arising from the interfaces \cite{tokuyasu88}, specifically when
the exchange splitting is strong, lead to different
current-phase relations for the spin-resolved currents $I_{\u}$ and $I_{\d}$
through the junction.  We show how this gives rise to
(i) a relative phase between single pair and ``crossed'' two-pair transmission; 
the latter process is
illustrated in
Fig.~\ref{fig1}a, with equal numbers of pairs transferred
in the spin $\u $ and spin $\d $ band;
(ii) different critical Josephson currents for opposite bias;
(iii) equilibrium shifts in the current phase relation,
in contrast to previous predictions \cite{Nazarov};
and (iv) a tunable spin supercurrent in a FM brought into contact with a
single SC electrode; we propose an experiment to
measure this remarkable effect.

%\textit{QUASICLASSICAL MODEL} -
Quasiclassical Green's functions (QCGF) \cite{serene,eilenberger}
are a powerful tool to describe hybrid structures of superconductors and non-superconducting materials. Consider, e.g.,
the interface between a SC and a sFM shown in Fig.~\ref{fig1}b.
For trajectories on the SC side, labeled 1, and characterized by Fermi momentum
$\vpfe $ and Fermi velocity $\vvfe $, the QCGF is obtained from the
microscopic one, $\hat G$, by integrating out the components oscillating on the Fermi wave length scale $\blFe =\hbar /\pfe $: $ \hat{g}(\vpfe , \vec{R}, \eps, t)=\int d\xi_p \hat{\tau}_3 \hat{G}(\vec{p}, \vec{R}, \eps, t)$,
where $\xi_p=\vvfe (\vec{p}-\vpfe )$.
The QCGF, $\hat{g}$, then varies as a function of the spatial coordinate $\vec{R}$ at a scale
set by the superconducting coherence length $\xi_0=\hbar \vfe /2\pi k_{\rm B} T_{\rm c}$,
and obeys the Eilenberger equation
\begin{equation}
\label{eilen} i \hbar \vvfe \cdot \nabla_{\vec{R}}\hat{g}+[\eps\hat{\tau}_3-\hat{\Delta}-\hat{h}, \hat{g}]=\hat{0},
\end{equation}
with normalization condition $\hat g^2 =-\pi^2 \hat 1$ \cite{eilenberger}.
Here, the hat denotes the $2\times2$ Nambu matrix structure in particle-hole space and $\hat{\tau}_3$ is the third Pauli matrix;
$\hat h$ includes all mean field and self energy terms governing
the quasiparticle motion along QC trajectories aligned with
$\vvfe $, and labeled by $\vpfe $;
$\hat \Delta $ is the SC order parameter.

The exchange field $\JFM $ in a sFM is comparable to the Fermi energy.
As opposed to the weak polarization
limit ($\JFM << \EF $) this cannot be described by a
term $-\vJFM \cdot \vec{\sigma }$ (with
$\vec{\sigma }$ the vector of Pauli spin matrices) in $\hat h$
of Eq.~(\ref{eilen}), because the QC approximation in this case
neglects terms of order $\JFM^2 /\EF $ compared to $\Delta $.
In most SCs, this is not justified for $\JFM >0.1\EF $.
However, for sufficiently large
$\JFM \gg \sqrt{\EF \Delta }$
the coherent coupling of the spin bands in the FM can be disregarded.
Consequently, we define an independent QCGF for each spin band
$\eta\in\{2,3\}$ in Fig.~\ref{fig1}b:
$
\hat{g}(\vpfeta , \vec{R}, \eps, t)=\int d\xi_{p\eta } \hat{\tau}_3\hat{G}(\vec{p}, \vec{R}, \eps, t),
$
where $\xi_{p\eta }=\vvfeta (\vec{p}-\vpfeta )$.
The exchange field is incorporated by the different Fermi velocities
$\vvfeta $ and momenta $\vpfeta $ in the two spin bands, and does not enter
the equation of motion~\eqref{eilen} for the QCGFs.
The $\hat{g}$ are Nambu matrices
with diagonal ($g$) and off-diagonal ($f$) components.
These components are {\it spin scalar}, as opposed to the QCGF in
the SC where they form a $2\times2$ spin matrix as a result of
spin coherence.
Indeed, the spins of the pair wavefunction in the
FM are fixed either to
$|\!\! \u\u\rangle$ (band 2 in Fig.~\ref{fig1}b) or to $|\!\! \d\d\rangle$ (band 3).

The interface enters the QC theory in the form of effective
boundary conditions \cite{Millis, Shelankov, Rica}
connecting the incident and outgoing QCGFs
for the three Fermi surface sheets $\eta \in \{1,2,3\}$.
The boundary conditions are subject to kinetic restrictions \cite{zutic00},
as illustrated in Fig.~\ref{fig1}b.
Note that, for a sFM, all singlet correlations are destroyed within the interface region (they decay on the short length scale
$\blJ=\hbar /(\pfz  -\pfd ) \ll \hbar \vfzd /\Delta \equiv \xi_{0\eta }$
\cite{condition}).
The boundary conditions are formulated in terms of the normal-state scattering matrix (S-matrix) of the interface \cite{Millis}, which for three bands
has the general form
\begin{equation}
\hat S=\hat{\varphi}
\left[\!\! \begin{array}{c|cc} \hat{R}_{11} & \vec{T}_{12} &
\vec{T}_{13} \\ \hline \vec{T}^T_{21} & r_{22} & r_{23}\\
\vec{T}^T_{31} & r_{32}  & r_{33}\end{array}
\!\!\right] \hat{\varphi}^{\dagger} ,\;
\hat{\varphi}=
\left[\!\!
\begin{array}{c|cc} e^{i\frac{\varphi}{2} \sigma_3}  & \!\! \vec{0} &\!\!
\vec{0} \\ \hline \vec{0}^T & e^{i\frac{\varphi}{2}} \!\! & \!\! 0\\
\vec{0}^T&\!\! 0  & \!\! e^{-i\frac{\varphi}{2}}\end{array} \!\!\right].
\label{eqS}
\end{equation}

\begin{figure}[t]
\includegraphics[width = 0.54\columnwidth]{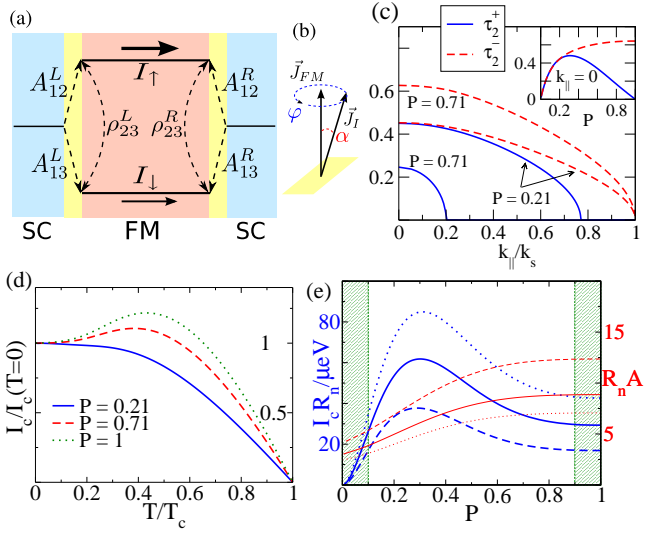}
\includegraphics[width = 0.42\columnwidth]{Figure2c.eps}
\\
\includegraphics[width = 0.45\columnwidth]{Figure2d.eps}
\includegraphics[width = 0.52\columnwidth]{Figure2e.eps}
\caption{\label{fig2}(color online) 
(a) Josephson-junction with spin-active SC/sFM interfaces  formed by
magnetized layers (yellow).
(b) Orientation of the interface magnetization
described by spherical angles $\alpha$ and $\varphi$.
(c) The quantities $\tau_2^{\pm}$ in Eq.~(\ref{tau}) vs.
$\vec{k}_{||}$ for two polarizations $P$, and vs. $P$
for perpendicular impact (inset).
(d) Critical current $I_{\mathrm{c}} $ vs. temperature $T$
for various polarizations $P$ of the sFM-layer.
(e) $I_{\mathrm{c}}R_{\mathrm{n}}$-product and normal state resistance $R_n
A$
as function of $P$ for
$T=0.5 T_\mathrm{c}$, $d=\ \blFe $, and
$(\VI-\JI )/\EF =10^{-4}$ (dotted), $0.2$ (full), $0.5$ (dashed).
$R_{\mathrm{n}}A$ in units of $(e^2 \Nfe \vfe )^{-1}$, $\Nfe $ being the
normal state SC DOS. $\Delta =1.76 $ meV.
In all plots: $\alpha_{\mathrm{L}}=\alpha_{\mathrm{R}}=\pi/2$,
$\phi_{\mathrm{L}}=\phi_{\mathrm{R}} $,
$L = \xi_0$, $d=5\blFe $, $\VI-\JI =0.5\ \EF $, $\pfz =1.18\ \pfe$, unless
stated otherwise. $P$ is tuned by $\pfd$.
}
\end{figure}

We obtain the reflection and transmission coefficients
from a microscopic calculation. We consider an
interface formed by a thin ($\approx \blF $) insulating FM layer of
thickness $d$ between the SC and bulk sFM (yellow areas in Figs.~\ref{fig1}b and
\ref{fig2}a),
characterized by an interface potential $\VI - \vJI \cdot \vec{\sigma }$.
The orientation of the
exchange field $\vJI $ in the interface layer
is determined by angles $\alpha$ and $\phi$, with $\alpha $
the angle between $\vJI $ and
the exchange field $\vJFM $ of the bulk sFM
(see Fig.~\ref{fig2}b).
The S-matrix connecting in- and outgoing amplitudes in the bulk SC and sFM is then obtained by a wave-matching technique, where the amplitudes in the interface layer are eliminated. Doing so, we obtain in the tunneling limit an S-matrix of the form $\hat{R}_{11}=e^{i(\vartheta/2)\sigma_3}$ \cite{Rs},
$\vec{T}_{12}=\vec{T}_{21}=
(t_{2}  e^{i\vartheta_{2}/2}, t'_{2} e^{-i\vartheta_{2}/2})^T$ and
$\vec{T}_{13}=\vec{T}_{31}= (t'_{3} e^{i\vartheta_{3}/2}, t_{3}
e^{-i\vartheta_{3}/2})^T$.
The spin mixing $\vartheta $-angles in these expressions
\cite{tokuyasu88,Millis,Eschrig03}
(also called spin-dependent interfacial phase shifts \cite{Brataas}), and
all remaining S-matrix parameters
are obtained from a microscopic calculation as outlined above.
As such, they depend on $d$, $\VI $,
$\alpha$ , $\phi$,
and the Fermi-momenta
of the three bands (we assume $|\vJI |=|\vJFM |$).
The dependence on the angle $\varphi$ is made explicit in Eq.~(\ref{eqS}),
while the dependence on the angle $\alpha$ is implicit in the
$r$ and $t$ parameters via
$t'_{2,3}\propto \sin(\alpha/2)$, $t_{2,3 }
\propto \cos(\alpha/2)$ and $\ r_{23},\ r_{32} \propto \sin\alpha$.
In the following we use these tunneling-limit expressions to gain
insight into the physics of the problem. The results shown in the figures
however, are obtained by a full numerical calculation.
For definiteness, we present results for parabolic electron bands with
equal effective masses.

%\textit{SC/FM/SC-JUNCTION} -
Applying these boundary conditions to a Josephson junction depicted in Fig.~\ref{fig2}a, and assuming bulk solutions for the QCGFs incoming from the SC electrodes, we arrive at the following system of linear equations for the $f$-functions in the tunneling limit (labels $k,j\in\{{\mathrm{L,R}}\}$ with $j\neq k$ denote the left/right interfaces):
\begin{equation}
\left[ \begin{array}{c} f_{2} \\ f_{3} \end{array}\right]^{out}_j \!\!  =
\left[ \begin{array}{cc} |r_{22}|^2 & \rho_{23}\\ \rho_{32}& |r_{33}|^2 \end{array}\right]_j
\left[ \begin{array}{c} \beta_{2} f_{2} \\ \beta_{3} f_{3} \end{array}\right]^{out}_k  \!\!+
\left[ \begin{array}{c} A_{12} \\ A_{13} \end{array}\right]_j.
\label{jbound}
\end{equation}
Here, the factors
$\beta_{\eta}=e^{-2 |\eps_n|L/v_{\perp\eta }}$,
where $L\gg \blJ $ is the junction length,
$\eps_n=(2n+1)\pi k_{\rm B}T$ the Matsubara frequency,
$v_{\perp \eta}$ the Fermi velocity component along the
interface normal,
and $\eta\in\{2,3\}$ the band index,
arise from the decay of the $f$ functions
in the sFM layer.
As depicted in Fig.~\ref{fig2}a,
coupling between the sFM spin bands is provided by the quantity
(for our model $r_{23}r^\ast_{32}$ is real)
\begin{equation}
\label{rho}
\left[ \rho_{23 }\right]_j =\left[r_{23}r^\ast_{32}e^{i2\varphi }\right]_j
=\left[\rho_{32}\right]_j^\ast,
\end{equation}
while the inhomogeneity in Eq.~(\ref{jbound}), $[A_{1\eta}]_j$, can be interpreted as a pair transmission amplitude from the SC into spin band $\eta$ of the sFM through the interface $j$. It reads
\begin{align}
\label{transamp}
&[A_{1\eta }]_j= -i\pi \; \frac{\mathrm{sgn}(\eps_n)}{1-\delta^2 }
\left[
(B_{\eta }+ C_{\eta })
t_{\eta }t'_{\eta  }\Delta e^{i(\chi \pm \varphi })
\right]_j\\
\label{B23}
&\qquad \qquad B_{\eta}=\tau_{\eta}^+/\Omega_n, \quad
C_{\eta }=\tau_{\eta}^-\cdot |\eps_n| /\Omega_n^2 \\
\label{tau}
&\tau_{\eta}^{\pm}=\sin\vartheta_{\eta }\pm\sin(\vartheta_{\eta}-\vartheta),
\quad \delta = \Delta \cdot \sin (\vartheta/2) /\Omega_n
\end{align}
where $\Omega_n=\sqrt{\eps_n^2+\Delta^2}$,
$\chi $ is the order parameter phase of the corresponding SC, and the $+(-)$ sign in $\chi\pm\varphi$ corresponds to $\eta =2(3)$.
Note that $t_{\eta}t_{\eta}'\propto \sin(\alpha)$, implying that the generation of triplet correlations relies on $\alpha \neq 0$.

In Fig.~\ref{fig2}c we show $\tau_\eta^\pm$, Eq.~(\ref{tau}),
for the majority spin band ($\eta=2$) as a function of $k_{||}$.
For large enough $k_{||}$, $\tau_2^+$ vanishes in contrast to $\tau_2^{-}$.
This region of $\vec{k}_{||}$-values allows for transmission into
only a single spin band of the FM (see Fig.~\ref{fig1}b).
With increasing spin polarization
%\begin{equation}
$
P=(\pfz -\pfd )/(\pfz +\pfd )
$
%\end{equation}
it extends over a larger range of $\vec{k}_{||}$-values, eventually spanning the entire Fermi-surface for a half-metal.
At the same time the maximal value of $\tau_2^+ $ decreases to zero, as
demonstrated in the inset in Fig.~\ref{fig2}c, where
the parameters $\tau_2^\pm $ are shown for normal impact as function of $P$.
The $\tau_\eta^{\pm}$ enter the $B_\eta $ and $C_\eta $ terms in
Eq.~(\ref{B23}), which exhibit different temperature ($T$) dependencies due to the
additional $|\eps_n|$ term in $C_\eta $ \cite{EschrigLTP}.
This interplay
leads to an intriguing change in the $T$-dependence of the Josephson current, plotted in Fig.~\ref{fig2}d.
For high $P$, a non-monotonic behavior is observed similar to that for a HM
\cite{Eschrig03,EschrigHM}, due to the dominant $C_2$ term,
whereas for smaller $P$
the term arising from $B_2$ leads to a monotonic decay with
increasing $T$. As a result, the bump in $I_{\mathrm{c}}(T)$ disappears with decreasing polarization.

\begin{figure}[t]
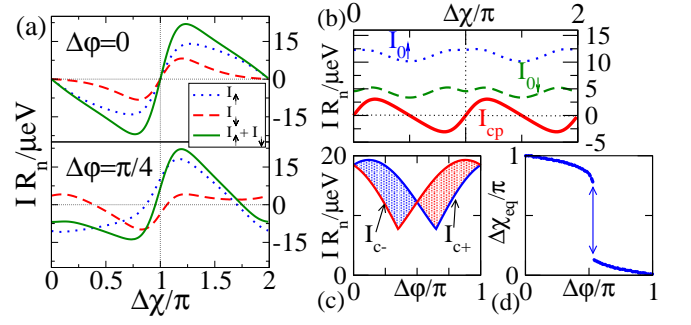

\includegraphics[width = 0.45\columnwidth]{Figure3a.eps}
\includegraphics[width = 0.53\columnwidth]{Figure3bcd.eps}
\caption{\label{fig3}
(color online)  
(a) Spin resolved and total CPR for $\Delta \varphi= 0 $ and $\pi/4$.
 (b) Coefficients $I_{\rm cp}$ and $I_{0\sigma }$ of Eq.~\eqref{CAR1}
vs. $\Delta \chi$.
 (c) Critical current in positive($I_{\rm c+}$)/negative($I_{\rm c-}$) bias direction vs. $\Delta \varphi $.
 (d) The
 equilibrium phase difference $\Delta\chi_{\rm{eq}}$
 vs. $\Delta\varphi$ varies from $\pi $ to 0.
 In all plots $T=0.2\ T_\mathrm{c}$, $d=0.25 \blFe $, $P=0.21$, other parameters as in Fig.~\ref{fig2}.
}
\end{figure}

In Fig.~\ref{fig2}e we plot
the $I_{\mathrm{c}} R_{\mathrm{n}}$ product as a function of $P$ (left scale).
The variation of the normal state resistance $R_n$ with $P$
(right scale)
cannot account for the variation of $I_{\mathrm{c}}$.
The critical current is suppressed
for small $P$ due to small spin mixing angles (see Fig.~\ref{fig2}c),
and for high $P$ due to reduction of conductivity in the minority spin band.
We thus predict a maximum critical current in a sFM junction for intermediate
$P\sim 0.3$. We caution that in the hatched regions in Fig.~\ref{fig2}e, there
are additional processes, not included in our model; e.g., for small $P$
spin coherence leads to singlet amplitudes in the FM.

We now discuss intriguing effects
associated with the angles
$\varphi_{\rm L,R} $ [see Eqs.~(\ref{rho}-\ref{transamp}) and
Fig.~\ref{fig2}b].
In Fig.~\ref{fig3}a we plot the spin-resolved
current-phase relation (CPR) \cite{golubov04} 
for a high transparency junction ($d=0.25\ \blFe $)
as a function of $\Delta\chi=\chi_{\mathrm{R}}-\chi_{\mathrm{L}}$ for two values of
$\Delta\varphi=\varphi_{\mathrm{R}}-\varphi_{\mathrm{L}}$
\cite{phi}. Clearly, there is a non-trivial modification of the CPR in the presence of $\Delta\varphi$.
We find that the CPR can be well described by the leading Fourier terms
in $\Delta \varphi$,
\begin{equation}
\label{CAR1}
I_{\sigma }\approx
I_{\rm cp}
-I_{0\sigma }\cdot \sin \Big(\Delta\chi_\sigma +\sigma \Delta\varphi  \Big)
\end{equation}
where $\sigma = +(-)1$ for spin $\u$($\d$). Here,
$I_{0\sigma } $ (shown in Fig.~\ref{fig3}b) and $\Delta\chi_\sigma $ are renormalized
due to multiple transmission processes.
The first term in Eq.~(\ref{CAR1}) describes a special type of
multiple transmission process, which we call ``crossed pair'' (cp) transmission, shown in Fig.~\ref{fig1}a.
It is a result of singlet-triplet mixing and triplet rotation induced by
the interfaces.
Here two singlet Cooper pairs are
effectively recombined coherently into two triplet pairs that propagate in different spin bands. Similar processes recombining a higher (but even) number of pairs will also contribute.
The phase associated with these processes comes from
$[A_{12}A_{13}]_{\rm L}[A_{12}A_{13}]_{\rm R}^*$ factors with $A_{1\eta}$ from Eq.~\eqref{transamp}, and is given by multiples of
$(\Delta\chi+\Delta\varphi)+(\Delta\chi-\Delta\phi)=2\Delta\chi$.
Consequently,
$I_{\rm cp} $ is independent of $\Delta \varphi$ and $\pi $-periodic in
$\Delta \chi$, as shown in Fig.~\ref{fig3}b (full line).
It is also obvious that $I_{\rm cp}$ is
spin symmetric, i.e., it carries a charge current, but no spin current.
We find that transfer processes with even number of pairs,
but non-zero total spin, are in contrast to the cp transmission
strongly suppressed.
Contributions to the second term in Eq.~(\ref{CAR1}) come from processes
that transmit one more Cooper pair in one of the spin bands compared to the
other, including single pair transmission.
It is therefore spin-dependent in magnitude (see Fig.~\ref{fig3}b) and
shows $\Delta \varphi $ phase shifts with opposite signs for opposite spins.
The relative phase between the two terms in Eq.~(\ref{CAR1}) leads to surprising measurable effects for finite $\Delta \phi$ and intermediate $P$.
First, we find a difference in the positive ($I_{\mathrm{c}+} $) and negative
($I_{\mathrm{c}-}$) bias critical charge currents, as shown in Fig.~\ref{fig3}c. This is also directly visible in Fig.~\ref{fig3}a, where the maximum and minimum current have a different absolute value.
Second,
we find a shift of the equilibrium phase $\Delta \chi_{\rm eq}$ for the
charge current, as shown in Fig.~\ref{fig3}d (the
jump as function of $\Delta \varphi$ is associated with multiple
local free energy minima).
We note that in the tunneling limit, Eq.~(\ref{CAR1}) reduces to
$ I_{\sigma }\approx -I_{\mathrm{0}\sigma }\cdot \sin(\Delta\chi+\sigma \Delta\varphi)$, and
the equilibrium phase shift is present as long as
$I_{\mathrm{0}\u}\ne I_{\mathrm{0}\d}$.

\begin{figure}[t]
\includegraphics[width = 0.35\columnwidth]{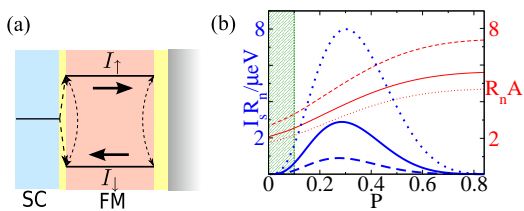}
\includegraphics[width = 0.43\columnwidth]{Figure4b.eps}
\caption{\label{fig4}(color online)  
(a) 
Setup with only one SC electrode.
(b) Spin-supercurrent $I_{\mathrm{s}}$ vs. $P$.
for various
$(\VI -\JI )/\EF =10^{-4}$(dotted), $0.2$(full), $0.5$(dashed).
$R_{\mathrm{n}}A$ refers to the normal state resistance of the SC/FM interface.
$d=\blFe $, other parameters as in Fig.~\ref{fig2}.}
\end{figure}

Another remarkable consequence of a non-zero $\Delta \varphi $ is observed
for a setup shown in Fig.~\ref{fig4}a, when an sFM is coupled
via a spin-active interlayer to a single SC on the left, and is terminated
by a magnetic surface on the right.
All quasiparticles are reflected at the
surface, leading to a zero charge current. However, not all
quasiparticles are reflected back into their original spin band since
spin-flip reflections [$\rho_{23}$ in Eq.~\eqref{rho}] mediate interactions
between the two spin bands, and, remarkably, a pure spin supercurrent remains. 
In this case, both terms in Eq.~(\ref{CAR1}) vanish as they are related to 
direct transmission. Instead, the leading term for the spin supercurrent is 
of second order in $\Delta \varphi$, $I \propto \sin(2\Delta \varphi )$, 
resulting from the phases picked up when a triplet Cooper pair reflects at the 
right interface \cite{brydon}. The 
maximal spin-current, defined  as
$I_{\mathrm{s}}=\mathrm{max}_{\Delta\varphi}I(\Delta\varphi)$, is plotted
in Fig.~\ref{fig4}b as a function of spin polarization.
Note that it vanishes both for $P\to 0$ and $P\to 1$,
since it requires the presence of two bands,
and is maximum for intermediate $P$.

This pure spin current can be tuned by an external microwave field that couples
to the magnetization of the right surface in Fig.~\ref{fig4}a,
and thus leads to a time dependent $\Delta \varphi (t)$.
A high degree of control can be achieved by manufacturing
a surface layer using a different magnetic material, preferably magnetized
perpendicular to the bulk FM, thus optimizing
external tunability.
As $\Delta \varphi (t)$ acts as a time dependent superconducting phase,
we predict in addition to a spin accumulation
in the FM a measurable ${\it ac}$ spin supercurrent,
analogously to the ${\it ac}$ charge Josephson current in a voltage biased junction.

%\textit{CONCLUSIONS}
In summary, we have presented a study of heterostructures between singlet
superconductors and strongly spin-polarized ferromagnets.
We have found that
the Josephson effect markedly differs from that for a
fully polarized material or for a ferromagnet with a weak spin band splitting.
We discussed the importance of the phase-shift between single pair and 
``crossed'' two-pair transfer processes that 
leads to measurable anomalous junction behavior.
We have also found
that a pure spin supercurrent is induced in a strongly polarized ferromagnet
coupled to one singlet superconducting electrode, and have proposed a way of measuring
this effect.

We thank T. L\"ofwander for stimulating discussions.

%\bibliography{scfmprl}

\begin{thebibliography}{11}
\bibitem{condition}
The condition for treating the singlet components within
QC theory is $\blJ^2 \gg \xi_{0\eta }\blFeta \gg \blFeta^2$.
\bibitem{Rs}
With a proper choice of the quantization axis in the SC, the reflection matrix $\hat{R}_{11}$ is spin-diagonal.
\bibitem{phi}
Our final results only depend on $\Delta \varphi=\varphi_{\mathrm{R}}-\varphi_{\mathrm{L}}$, in
accordance with global spin rotational invariance.
\end{thebibliography}

\end{document}